\documentclass[11pt,twoside]{article}

\usepackage{asp2006}
\usepackage{epsf}
\usepackage{lscape}
\usepackage{graphicx}

\begin{document}
\title{Interacting Galaxies with MOND}   
\author{O. Tiret and F. Combes}   
\affil{Observatoire de Paris, LERMA}    

\begin{abstract} 
We compare N-body simulations performed in MOND with analogs in Newtonian gravity with dark matter (DM). We have developed a code which solves the Poisson equation in both gravity models. It is a grid solver using adaptive mesh refinement techniques, allowing us to study isolated galaxies as well as interacting galaxies. Galaxies in MOND are found to form bars faster and stronger than in the DM model. In Newton dynamics, it is difficult to reproduce the observed high frequency of strong bars, while MOND appears to fit better the observations. Galaxy interactions and mergers, such as the Antennae, are also simulated with Newton and MOND dynamics. In the latter, dynamical friction is much weaker, and merging time-scales are longer. The formation of tidal dwarf galaxies in tidal tails are also compared in MOND and Newton+DM models.
\end{abstract}


\section{Introduction}   
To approach the missing mass problem, two ways are nowadays seriously considered. The first makes the hypothesis that Newtonian gravity or Einstein's General Relativity is the good theory of Gravitation, which leads to introduce dark matter: it is the $\Lambda$CDM model. The second consists in considering only the existence of baryonic matter, implying a breaking of the usual gravitation law, which is called MOND (MOdified Newtonian Dynamics).
In highly symmetric geometries (spherical for example), the modification of the gravity can be written like this (Milgrom, 1983):

$$a_{Newt}= a \mu (a/a_0),$$

\noindent where $a_0\sim 1.2\times10^{-10}m.s^{-2}$ is the critical acceleration of MOND, and $\mu$ an interpolating function between the MONDian and Newtonian regimes, which has the following properties:

$$\mu(x) \left\{
\begin{array}{l}
  1:x \gg 1\\
  x:x \ll 1
\end{array}
\right.$$

\noindent Through this formulation of the gravitation, MOND has the ability to fit a large number of rotation curves, fixing the value of $a_0$ and choosing the same $\mu$-function whatever the galaxy (Sanders \& McGaugh, 2002).

\newpage
We propose to test MOND on galaxy dynamics using N-body simulations. To do that we had to develop a solver for the MOND Poisson equation (Bekenstein \& Milgrom, 1984):

$$\overrightarrow{\bigtriangledown} \cdot \left[ \mu\left( \frac{\mid \overrightarrow{\bigtriangledown} \phi \mid}{a_0} \right) \overrightarrow{\bigtriangledown} \phi \right]=4\pi G \rho.$$ 

\noindent It is a grid solver using multigrid techniques. The code is described in Tiret \& Combes (2007), hereafter TC07. Each simulation is run in the MOND model and its analogue in Newtonian gravity with DM.

\section{Isolated Galaxy}

In the DM model, the galaxy disc feels the gravitational potential from the disc itself, plus from the dark matter halo. Inside the disc, the stars are dominated by ordered motions, in nearly circular rotation, thus the disc is sensitive to gravitational instabilities. On the contrary, the DM particles in the halo are dominated by random motions, which maintain the halo stable and spheroidal. The contribution of the halo is then to produce a stable background potential,
and hence to stabilize the galaxy disc, in reducing its self-gravity.

\begin{figure}[b]
 \centering
 \includegraphics[width=10 cm]{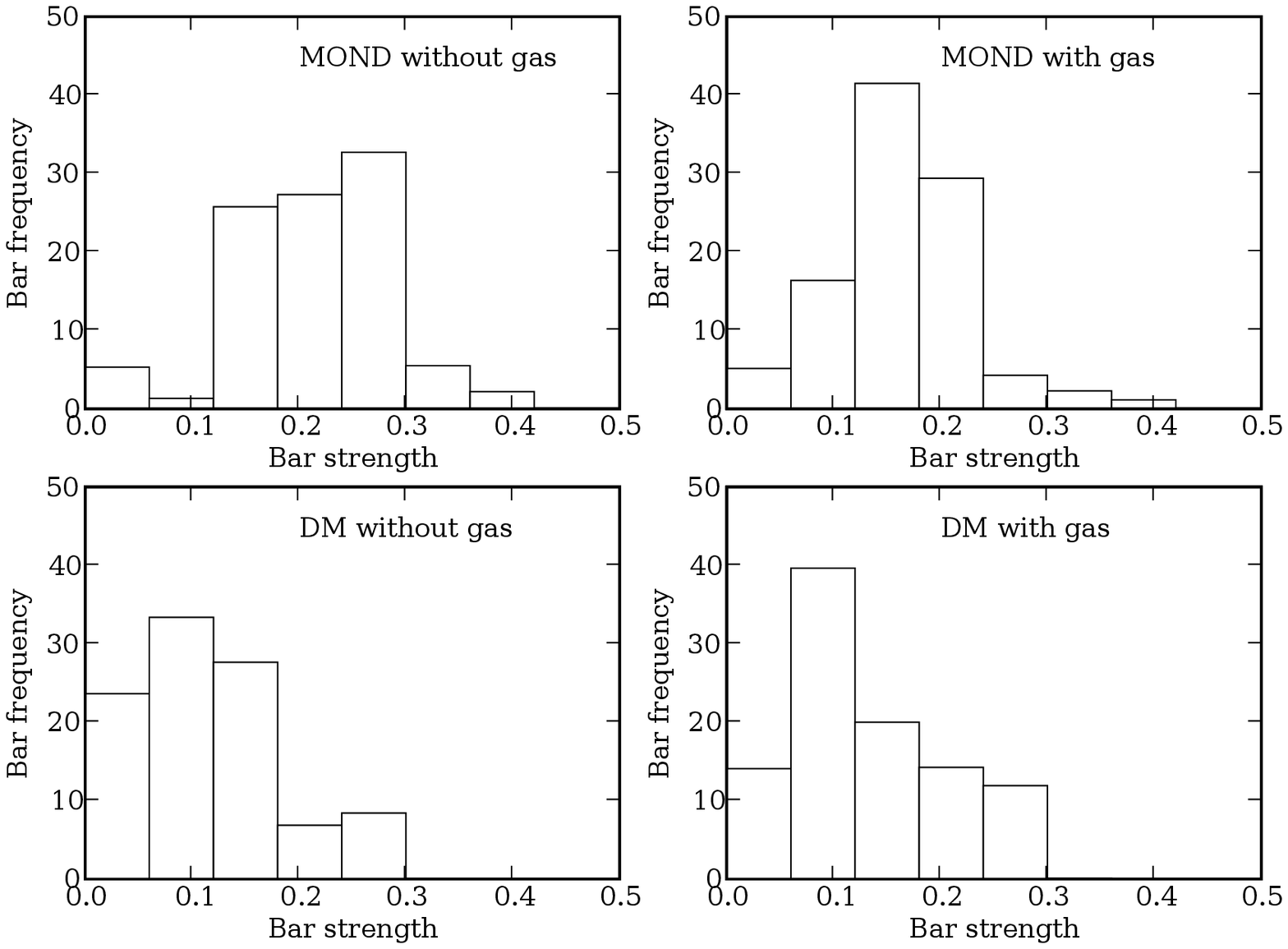}
 \caption{Histograms of bar frequency in a simulated Hubble sequence, with MOND (top) and DM (bottom), and with gas (right) and without (left).}
 \label{fig:bar_freq}
\end{figure}

In MOND, the potential comes only from the disc, so that the galaxy disc is completely self-gravitating. In pure stellar simulations (TC07), we found that galaxies in MOND are more unstable in the sense that they form bars stronger and faster than in the analog DM model. New simulations have been carried out, adding a cold gas disc (Tiret \& Combes in prep.). In MOND, it does not change significantly the stability of galaxies, except for late-types where the fraction of gas is about $7\%$ inside the visible radius, sufficient to destroy bars by the effect of gravity torques (Bournaud et al., 2005). Otherwise, the introduction of gas affects more the DM model, where bar strengths were not yet saturated. Galaxies are more unstable and form bars sooner. Diagrams of bar frequency are then comparable in the MOND and DM models (Fig. \ref{fig:bar_freq}).

\section{Galaxy interactions}
\begin{figure}
 \centering
 \includegraphics[width=\textwidth]{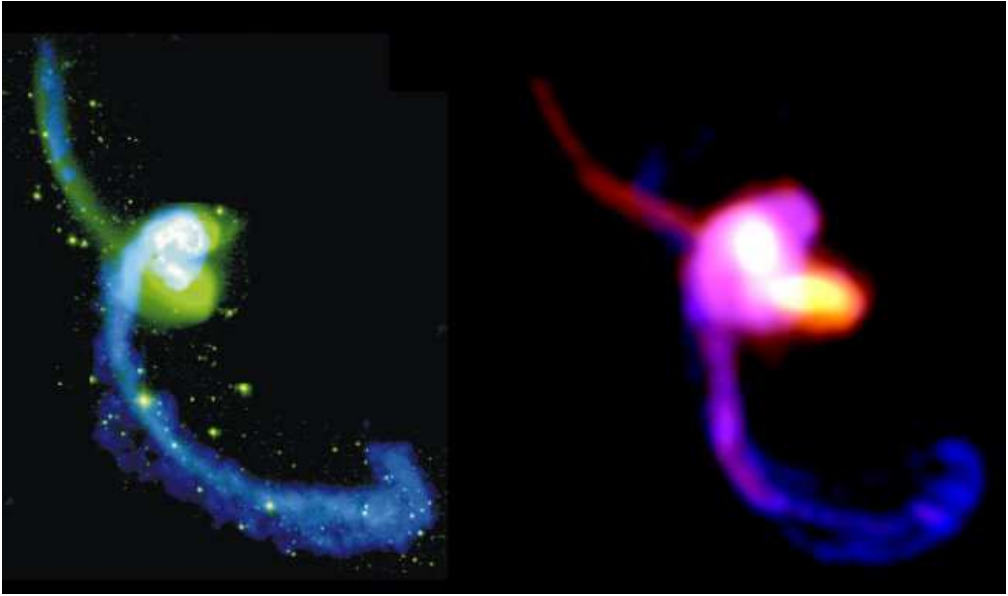}
 \caption{Simulation of The Antennae with MOND (right) compared to the observations from Hibbard et al. 2001 (left).
In the observations, the gas is represented in blue and the stars in green. In the simulation the gas is in blue and the stars are in yellow/red.}
 \label{fig:ant_morpho}
\end{figure}

It is essential to test MOND on scales where it has not been built for. After studying isolated galaxies, the next step is to apply MOND on interacting galaxies. We are interested to simulate a system like the prototypical Antennae Galaxies. The observations (Hibbard et al., 2001) show quite extended tidal tails. This system has been simulated many times in Newtonian gravity (Barnes, 1988, or Dubinski et al. 1996) where the DM halo mass and shape were constrained by the length of the tidal tails. The question here is to know whether these extended tidal tails could also form with MOND. Using our code in its adaptive resolution version, we run the simulation, and show that the morphology of the Antennae galaxies is well reproduced with MOND (Fig. \ref{fig:ant_morpho}). We compare also the kinematics of the simulated system to the observations (Tiret \& Combes in prep.).
 
From one point, the two models differ radically: the dynamical friction. In the DM model, the two galaxies quickly lose their orbital momentum which is transferred to the dark halo, the time-scale to obtain the merger is of the order of a few gigayears.
In MOND, the dynamical friction which tends to decrease the distance between the galaxies until the coalescence, is much weaker, in the absence of extended massive haloes. All the orbital momentum is converted in angular momentum of the outer discs. The time-scale of the merger could be a few gigayears if the impact parameter is about equal to the visible radius of the galaxies. Otherwise, the galaxies decay without merging in a Hubble time (see also Nipoti et al. 2007).

We also show that tidal dwarf galaxies can be naturally formed at the tip of the tidal tails, in MOND. While it requires extended dark matter halo in a DM model (Bournaud et al. 2003).

\section{Conclusion}
Numerical simulations offer a powerful tool to test galaxy dynamics with MOND. We have developed an efficient potential solver for the modified Poisson equation. With this, the stability of galaxy discs can be studied and compared to Newtonian gravity with DM. Simulation of interacting galaxies have also been performed. We show that MOND can reproduce the morphology as well as the kinematics of The Antennae Galaxies. At this stage, we are not able to discriminate between the two models.
But some points have to be explored, in particular the fact that mergers in MOND are more difficult to obtain because of the weaker dynamical friction. So we expect large differences in the cosmic merger history between the MOND and DM model. However it is possible that we observe the same proportion of galaxies showing signs of interaction whatever the model considered. In the DM model, the duration of each merger is short, but their number is high, while in MOND the number of mergers is small but each lasts a longer time. A possible test is to determine observationally the number of mergers that galaxies experience during their life time, still an uncertain quantity.


\end{document}